\input{aipcheck}
\newcommand{\bv}[1]{{\mathbf #1}}
\documentclass[
    ,final            % use final for the camera ready runs
letterpaper
  ]
  {aipproc}

\layoutstyle{8x11single}
\begin{document}

\title{Long-range Correlation in Sheared Granular Fluids}


\author{Michio Otsuki}{
  address={Yukawa Institute for Theoretical Physics, Kyoto University,  Kitashirakawa-oiwake cho, Sakyo,
Kyoto 606-8502, JAPAN}
}

\author{Hisao Hayakawa}{
}

\begin{abstract}

The spatial correlation function of the momentum density
in the three-dimensional dilute sheared granular fluids is theoretically
investigated. 
The existence of the long-range
correlation is verified through both analytic
calculation and numerical simulation.
\end{abstract}

\maketitle

%%%%%%%%%%%%%%%%%%%%%%%%%%%%%%%%%%%%%%%%%%%%
%% MAINMATTER
%%%%%%%%%%%%%%%%%%%%%%%%%%%%%%%%%%%%%%%%%%%%
\section{Introduction}

% review of the time and spatial correlation

The time and the spatial correlations play important
roles in non-equilibrium statistical physics [1-16].
The behaviors of the correlation functions in ordinary fluids 
are well understood.
It is known that there exist long-time tails in the
time correlation functions in fluids at equilibrium
 \cite{alder,Pomeau,kawasaki-oppenheim,ernst71}. 
In addition, long-range correlations exist
in the spatial correlation functions in non-equilibrium ordinary fluids 
 \cite{Dorfman,Machta,Wada,Lutsko}.

% time correlation in granular fluids

On the other hand, interest in 
the correlations of granular fluids is rapidly growing.
In the case of the freely cooling state, 
it is confirmed that
long-time tails  for
the time-correlation of the velocity and the shear stress 
exist but the time-correlation function
for the heat flux decays exponentially \cite{Hayakawa}.
In addition, the spatial correlation
functions are analytically calculated by using fluctuating hydrodynamics
and the theoretical results are verified by numerical simulations \cite{Noije}.
See also the studies on the long-time tails in granular flows \cite{Orpe,Hayakawa2}. 
In sheared granular fluids, there are some studies on
time correlations, 
but the situation is still confusing \cite{Kumaran, Otsuki,Otsuki2}.
Indeed, Kumaran predicted that the correlation function satisfies $t^{-3d/2}$
with the spatial dimension $d$ \cite{Kumaran},
while we obtained crossover from $t^{-d/2}$ to $t^{-(d+2)/2}$ for
the velocity auto-correlation function of nearly elastic granular gases \cite{Otsuki}.
However, there is no corresponding theoretical or numerical argument
on the spatial correlation functions in sheared
granular fluids.

% spatial correlation in granular fluids
In this paper, thus,  we investigate the spatial correlation functions
in sheared granular fluids.  
In section 2.1, we will explain the set up.  In section 2.2, we will present  
the analytic results for the spatial correlation of the momentum density.
In section 2.3, the validity of the analytic results will be tested by our numerical simulation.
Finally, we will discuss and conclude our results in section 3.

\section{Result}

\subsection{Set up}

% hard spherical particles
Let us consider
a three-dimensional dilute system consisting of $N$ identical smooth and hard spherical particles with mass
$m$ and 
diameter $\sigma$ in the volume $V$. 
The position and the velocity of the $i$-th particle at time $t$
are denoted by $\bv{r}_i(t)$ and $\bv{v}_i(t)$, respectively.  
The particles collide instantaneously with each other
with a restitution constant $e$ which is less than unity for granular particles.
When the particle $i$ with velocity $\bv{v}_i$ collides with the particle  $j$ with $\bv{v}_j$,
the post-collisional velocities $\bv{v}'_i$ and $\bv{v}'_j$ are respectively given by
%\begin{eqnarray}\label{eq1}
$\bv{v}'_i  =  \bv{v}_i  - \frac{1}{2} (1+e) (\bv{n} \cdot
\bv{v}_{ij}) \bv{n} $ and
$\bv{v}'_j  =  \bv{v}_j  + \frac{1}{2} (1+e) (\bv{n} \cdot
\bv{v}_{ij})\bv{n} $,
%\end{eqnarray}
where $\bv{n}$ is the unit vector parallel to the relative position of the
two colliding particles at contact, and $\bv{v}_{ij} \equiv \bv{v}_{i}-\bv{v}_{j}$. 
Let us assume that the uniform shear flow is stable and
%analyze the velocity autocorrelation function under the shear 
%flow in which the average velocity profile is given as 
its velocity profile is given by
$c_{\alpha}(\bv{r}) = \dot{\gamma} y \delta_{\alpha,x}$,
where the Greek suffix $\alpha$ denotes the Cartesian component,
and $\dot{\gamma}$ is the shear rate.

% observable

Let us consider the spatial correlation function 
of the momentum density defined by
\begin{eqnarray}
C_{pp}(\bv{r}) \equiv \int \frac{d \bv{r}'}{V} 
 \left < \left [ \bv{p}(\bv{r}+\bv{r}',t) - \rho(\bv{r}+\bv{r}',t)\bv{c}(\bv{r}+\bv{r}') \right ] 
\cdot  \left [ \bv{p}(\bv{r}',t) - \rho(\bv{r}',t)\bv{c}(\bv{r}') \right ]\right >,
\end{eqnarray}
where $\bv{p}(\bv{r},t) \equiv m \sum_i^N \bv{v}_i(t) \delta(\bv{r}-\bv{r}_i(t))$ and
$\rho(\bv{r},t)\equiv m \sum_i^N \delta(\bv{r}-\bv{r}_i(t))$ are
the momentum density and the density, respectively.

\subsection{Theoretical Analysis}

% fluctuating hydrodynamics
In order to obtain the analytic expression of $C_{pp}(\bv{r})$, 
we assume that the time evolution of the hydrodynamic fields
is described by fluctuating hydrodynamics \cite{Landau}
\begin{eqnarray}
\partial_t \rho + \bv{\nabla} \cdot (\rho\bv{u}) & = & 0, \label{n:eq} \\
\partial_t \bv{u} + \bv{u} \cdot \bv{\nabla} \bv{u} 
+ \bv{\nabla} \cdot \Pi/ \rho
& = & 0,  \label{u:eq} \\
\partial_t T + \bv{u} \cdot \bv{\nabla} T 
+ 2 m(\Pi:\bv{\nabla}\bv{u} 
+  \bv{\nabla} \cdot \bv{q}) /(3\rho)
& = & -\zeta T , \label{T:eq}
\end{eqnarray}
where  $\bv{u}(\bv{r},t)\equiv \bv{p}(\bv{r},t)/\rho(\bv{r},t)$ and $T(\bv{r},t)$ 
are  the velocity and the temperature, respectively.
The heat flux $\bv{q}$ and the pressure tensor $\Pi_{ij}$
consist of two parts as $\bv{q}=\bv{q}^*+\bv{q}^R$ and 
$\Pi_{ij} = \Pi_{ij}^* + \Pi_{ij}^R$.
Here, $\bv{q}^*$ and
$\Pi_{ij}^*$ represent systematic parts as
$\bv{q}^* = 
-\kappa \bv{\nabla}  T - \mu \bv{\nabla} \rho/m$, and
$\Pi_{ij}^* = \rho T \delta_{ij}/m - \eta  [\nabla_i u_j + \nabla_j u_i  -(2 \eta /3) \delta_{ij} \nabla_k u_k]$, respectively, where $\delta_{ij}$ is Kronecker delta.
Note that the bulk viscosity disappears in fluids of dilute spherical
particles.
$\zeta$, $\kappa$, $\mu$, and $\eta$
are the cooling rate,
the heat conductivity, 
the transport coefficient associated with the
density gradient, and the viscosity, respectively.
Here, $\mu$ has a finite value when $e$ is less than unity \cite{Brey}.
$\bv{q}^R$ and $\Pi_{ij}^R$ are the random parts of the 
heat flux and the pressure tensor,
respectively.
$\bv{q}^R$ and $\Pi_{ij}^R$ are respectively written as
$q_i^R =  \sqrt{T^2 \lambda } f_i^h$, and 
 $\Pi_{ij}^R = \sqrt{T \eta} f_{ij}^s$,
 where $f_i^h$ and $f_{ij}^s$ satisfy
$\left < f_i^h \right >  =  \left <  f_{ij}^s \right >  = 0$,
$\left < f_i^h f_{ij}^s  \right >  = 0$,
$\left < f_i^s (\bv{r},t) f_j^s (\bv{r}',t')\right >  = 
2\delta_{ij} \delta(\bv{r}-\bv{r}') \delta(t-t')$, and
$\left < f^s_{ij}(\bv{r},t) f^s_{kl}(\bv{r}',t')\right >  
 =  2\Delta_{ijkl} \delta(\bv{r}-\bv{r}') \delta(t-t')$
 with
$\Delta_{ijkl} = \delta_{ik} \delta_{jl} + \delta_{il} \delta_{jk}
- 2 \delta_{ij} \delta_{kl} /3$.

Let the viscosity $\eta$, the cooling rate $\zeta$, the heat conductivity $\kappa$, and the transport coefficient associated with the density gradient $\mu$
be non-dimensionalized as 
\begin{equation}
\eta  =  \eta_0 \eta^*, \quad
\zeta = \rho T \zeta^* / \eta_0, \quad
\kappa  =  \kappa_0 \kappa^*, \quad
\mu  =  m T\kappa_0 \mu^*/\rho, \label{eta}
\end{equation}
where $\eta^*$, $\zeta^*$, $\kappa^*$, and $\mu^*$ are constants 
which depend only on $e$ in dilute cases \cite{Brey}. 
We note  $\eta^* \simeq 1$ and $\zeta^* \simeq 5 \epsilon /12$
with $\epsilon = 1 - e^2$ in the limit of small $\epsilon$.
Here, $\eta_0$ and $\kappa_0$ 
are the viscosity and the heat conductivity in the dilute elastic hard-core gas 
given by
\begin{equation}
\eta_0  =  a \sqrt{T}, \quad
\kappa_0  =  15 a \sqrt{T} / (4m), \label{eta0}
\end{equation}
 respectively.
The explicit form of the constant $a$ is given by
$a = 5 \sqrt{m/\pi} 
/ (16 \sigma^{2})$.

% TH

Here, we introduce the average density 
$\rho_0 \equiv \left < \rho(\bv{r},t) \right >$, and the average temperature
$T_0 \equiv \left < T(\bv{r},t) \right >$.
These averages satisfy $\rho_0 = mN/V$, and  
  \begin{equation}
T_{0} =  2m^2 a^2 \eta^* \dot{\gamma}^2/(3\rho_0^2 \zeta^*). \label{T_0}
\end{equation}

% Method

The method to calculate the spatial correlation function is parallel to
that for the sheared ordinary fluids \cite{Lutsko}. We, thus, 
show only the outline of the method in this paper.
First, we introduce the fluctuations of the hydrodynamic fields
$\delta \rho(\bv{r},t) \equiv \rho(\bv{r},t) - \rho_0$,
 $\delta T(\bv{r},t) \equiv T(\bv{r},t) - T_0$, and
  $\delta \bv{u}(\bv{r},t) \equiv \bv{u}(\bv{r},t) - \bv{c}(\bv{r},t)$.
Then, $C_{pp}(\bv{r})$ can be approximated by
\begin{eqnarray}
C_{pp}(\bv{r}) & = & \int \frac{d \bv{r}'}{V} 
 \left <  \{\rho_0 + \delta \rho(\bv{r}+\bv{r}',t)\} \delta \bv{u}(\bv{r}+\bv{r}',t)
\cdot  \{\rho_0 + \delta \rho(\bv{r}',t) \} \delta \bv{u}(\bv{r}',t)\right > \nonumber \\ 
& \simeq & \rho_0^2 \int \frac{d \bv{r}'}{V} 
 \left <   \delta \bv{u}(\bv{r}+\bv{r}',t)
\cdot  \delta \bv{u}(\bv{r}',t)\right > 
 =  \rho_0^2 \int \frac{d \bv{k}}{(2\pi)^3} \tilde{C}_{uu}(\bv{k}) 
e^{i\bv{k} \cdot \bv{r}}, \label{Cp:app}
\end{eqnarray}
where we have ignored nonlinear terms of the fluctuations.
We have also
introduced the correlation function $\tilde{C}_{uu}(\bv{k})$, which satisfies
\begin{equation}
\left < \delta\bv{u}_{\bv{k}}(t) \cdot 
\delta\bv{u}_{\bv{k}'}(t)\right >= (2\pi)^3 \delta^3(\bv{k}+\bv{k}') \tilde{C}_{uu}(\bv{k}).
\end{equation}

Let us introduce the vector $\bv{z}(\bv{r},t)$ and its Fourier transform of the fluctuations as
\begin{equation}
\tilde{\bv{z}}^T(\bv{k},t)= (\delta n_{\bv{k}}(t), 
\delta T_{\bv{k}}(t), \delta u^{(1)}_{\bv{k}}(t), \delta u^{(2)}_{\bv{k}}(t), \delta u^{(3)}_\bv{k}(t)), \label{z:def}
\end{equation}
where we decompose $\delta \bv{u}_{\bv{k}}(t)$ 
 as $\delta \bv{u} _{\bv{k}}(t) = \delta u^{(1)}_\bv{k}(t) \bv{e}^{(1)} + 
  \delta u^{(2)}_{\bv{k}}(t) \bv{e}^{(2)} + \delta u^{(3)}_\bv{k}(t) \bv{e}^{(3)}$
 with 
 $\bv{e}^{(1)}\equiv \bv{k}/k$,
 $\bv{e}^{(2)}\equiv \{ \bv{e}_{y} 
- (\bv{e}^{(1)} \cdot \bv{e}_{y}) \bv{e}^{(1)} \} / {\cal N}$,
$\bv{e}^{(3)} \equiv \bv{e}^{(1)} \times \bv{e}^{(2)}$,
and $\delta u^{(i)}_\bv{k}(t) = \delta \bv{u}_{\bv{k}}(t) \cdot \bv{e}^{(i)}$.
Here, $\bv{e}_y = (0,1,0)$, and
${\cal N} = |  \bv{e}_{y} 
- (\bv{e}^{(1)} \cdot \bv{e}_{y}) \bv{e}^{(1)}|$.
Then,  the linearized time evolution equation for $\tilde{\bv{z}}(\bv{k},t)$ is obtained
from eqs. (\ref{n:eq}), (\ref{u:eq}), (\ref{T:eq}), and (\ref{z:def}) as
\begin{equation}
(\partial_t - \dot{\gamma} k_x \partial_{k_y}) \tilde{z}_{\alpha} (\bv{k},t)
+ L_{\alpha \beta} \tilde{z}_{\beta} (\bv{k},t) = \tilde{R}_\alpha(\bv{k},t).
 \label{Lin:eq}
\end{equation}
Here, the matrix $L_{\alpha \beta}$ is given by
\begin{equation}
L_{\alpha \beta} = ik L^{(1)}_{\alpha \beta} + k^2 L^{(2)}_{\alpha \beta}
 + \dot{\gamma} L^{(3)}_{\alpha \beta}
  + \dot{\gamma} ik L^{(4)}_{\alpha \beta}
  + \dot{\gamma}^2 L^{(5)}_{\alpha \beta}, \label{matrix}
  \end{equation}
where $L^{(i)}_{\alpha \beta}$ is a matrix depending on
$\zeta$, $\kappa$, $\mu$, and $\eta$.
$\tilde{R}_\alpha(\bv{k},t)$ is a random vector
which is a function of $\bv{q}^R$ and $\Pi_{ij}^R$.
See the elements of the matrix $L^{(i)}_{\alpha \beta}$
and the vector $\tilde{R}_\alpha(\bv{k},t)$ in Appendix A.

The solution of eq. (\ref{Lin:eq}) is expressed as 
\begin{equation}
\tilde{z}_{\alpha} (\bv{k},t) = \sum_i^5 \int _{-\infty}^{t} ds
\psi^{(i)}_\alpha(\bv{k},t-s) F^{(i)}(\bv{k}(\dot{\gamma}(s-t)),s), \label{z:sol}
\end{equation}
where 
\begin{equation}
\psi^{(i)}_\alpha(\bv{k},t) \equiv 
\psi^{(i)}_\alpha(\bv{k}) e^ { - \int_0^t ds \lambda^{(i)}( \bv{k}(\dot{\gamma}s))}, \qquad
F^{(i)}(\bv{k},t) \equiv \phi^{(i)}_\beta(\bv{k})\tilde{R}_\alpha(\bv{k},t)
\end{equation}
with $\bv{k}(\tau) \equiv (k_x,k_y - \tau k_x,k_z)$.
Here, we have introduced the linearly independent eigenvectors 
$\psi^{(i)}_\alpha(\bv{k})$, 
the associated biorthogonal vectors $\phi^{(i)}_\alpha(\bv{k})$, 
and the eigenvalues $\lambda^{(i)}(\bv{k})$ satisfying
\begin{equation}
(- \dot{\gamma} k_x \partial_{k_y} \delta_{\alpha \beta} + L_{\alpha \beta})
\psi^{(i)}_\beta(\bv{k}) = \lambda^{(i)}(\bv{k}) \psi^{(i)}_\beta(\bv{k}),
\end{equation}
and $\psi^{(i)}_\alpha(\bv{k}) \phi^{(j)}_\alpha(\bv{k})= \delta_{ij}$.

In order to obtain an analytic expression of
$\psi^{(i)}_\alpha(\bv{k})$, 
$\phi^{(i)}_\alpha(\bv{k})$, 
and $\lambda^{(i)}(\bv{k})$,
we assume  \cite{Lutsko}
\begin{equation}
\dot{\gamma} \sim O(\eta k^2 / \rho). \label{condition}
\end{equation}
Here, we note that eq. (\ref{condition}) is rewritten as
$l_c^{-2} \sim O(k^2)$,
where $l_c$ is the length scale defined by
\begin{eqnarray}
l_c \equiv \sqrt{\frac{2 \eta}{m n_0 \dot{\gamma} }}. 
\label{threshold}
\end{eqnarray}
From eqs. (\ref{eta}), (\ref{eta0}) and (\ref{T_0}),
the length scale $l_c$ becomes
\begin{eqnarray}
l_c = \frac{5 \sqrt{\pi}}{96} \nu^{-1}  
\left (  \frac{2\eta^{*3}}{3\zeta^*} \right )^{1/4} \sigma,
\end{eqnarray}
where $\nu$ is the volume fraction.
We, thus, find that $l_c$ does not depend on $\dot{\gamma}$, but depends on the volume fraction $\nu$ and 
the restitution constant $e$ because $\eta^*$ and $\zeta^*$
depend only on $e$.
In the limit of small $\epsilon$, we obtain 
$l_c \propto \sigma \nu^{-1} \epsilon^{-1/4}$.

Substituting the solution (\ref{z:sol})  for small $k$
into eq. (\ref{Cp:app}),
we obtain an analytic expression of $C_{pp}(\bv{r})$ as
\begin{eqnarray} 
C_{pp}(\bv{r}) =  \frac{ \rho_0 T_0}{l_c^3} 
\left \{ 
  \Delta_1(\bv{r}/l_c) +  \Delta_2(\bv{r}/l_c) +  \Delta_3(\bv{r}/l_c) 
  \right \}, \label{Result}
\end{eqnarray}
where
\begin{eqnarray}
\Delta_1(\tilde{\bv{r}}) 
& = & \int \frac{d \bv{k}}{(2 \pi)^3} e^{-i \bv{k} \cdot \tilde{\bv{r}}}
\int_0^{\infty} dt k k(t) [ a_1 - a_2\cos(c_1 k \alpha(t))] 
e^{- b  (t k^2 +t^2k_x k_y + t^3 k_x^2/3) } , \nonumber \\
\Delta_2(\tilde{\bv{r}}) 
& = & \int \frac{d \bv{k}}{(2 \pi)^3}  e^{-i \bv{k} \cdot \tilde{\bv{r}}}
\int_0^{\infty} dt \frac{k(t)^4}{k^2}  
e^{-   (t k^2 +t^2k_x k_y + t^3 k_x^2/3) }, \nonumber \\
\Delta_3(\tilde{\bv{r}}) 
& = & \int \frac{d \bv{k}}{(2 \pi)^3}  e^{-i \bv{k} \cdot \tilde{\bv{r}}}
\int_0^{\infty} dt  \frac{k(t)^2}{k^2}  
e^{-   (t k^2 +t^2k_x k_y + t^3 k_x^2/3) }\nonumber \\
& & \times 
[ M(\bv{k}(t))^2 k^2 - 2 M(\bv{k}(t))M(\bv{k}) k\bv{k}(t) 
+ M(\bv{k})^2 k(t)^2 + k(t)^2
], \label{Delta:def}
\end{eqnarray}
where $\bar{\bv{r}} \equiv \bv{r}/l_c$.
Here, 
$M(\bv{k}) = - k k_z / (k_x \sqrt{k^2-k_y^2}) \tan^{-1}(k_y / \sqrt{k^2-k_y^2})$,
$a_1 = (2\eta^*/3 + \kappa^*/2 )/\eta^*$,
$a_2 = (  \kappa^*/2 -2  \eta^* /3)/\eta^*$,
$b = (2\eta^*/3 + \kappa^*/2 + 3\mu^*/4 )/\eta^*$, and 
$c_1 = 2 \sqrt{5 \nu_{0}}/ \sqrt{3 \dot{\gamma} \eta^*}$
with $\nu_{0} = \rho_0 T_0/(m\eta_0(T_0))$.
It should be noted that the expressions in eqs. (\ref{Result}) and (\ref{Delta:def})
are the same as those for the sheared ordinary fluids by Lutsko and Dufty \cite{Lutsko}.
The difference between ours and theirs exists in the dependence on the restitution constant $e$ through $l_c$.

Let us  explicitly demonstrate the existence of the long-range correlation in
$C_{pp}(\bv{r})$.
Let the angular average of any function $f(\bv{r})$ be denoted by
$\bar{f}(r) \equiv  \int d \Omega / (4 \pi) f(\bv{r})$.
From the second equation in eq. (\ref{Delta:def}), it is easy to show 
the asymptotic behavior 
\begin{eqnarray}
\bar{\Delta}_2(\tilde{r})  \propto \tilde{r}^{-5/3}, \qquad \tilde{r} \gg 1.
\label{Delta2:long}
\end{eqnarray}
Indeed, from the transformation of variables as
$k' = k\tilde{r}$ and $s = t \tilde{r}^{-2/3}$,
we obtain
\begin{eqnarray}
\bar{\Delta}_2(\tilde{r}) 
& = & \int \frac{d \bv{k}}{(2 \pi)^3} \frac{sin(k\tilde{r})}{k\tilde{r}}
\int_0^{\infty} dt \frac{(k^2+2t k_x k_y + t^2 k_x^2)^2}{k^2}  
e^{-  (t k^2 +t^2k_x k_y + t^3 k_x^2/3) } \nonumber \\
& = & 
\tilde{r}^{-5/3} \int \frac{d \bv{k}'}{(2 \pi)^3} \frac{sin(k')}{k'}
\int_0^{\infty} ds \frac{(\tilde{r}^{-4/3} k^{'2}+2\tilde{r}^{-2/3}s k'_x k'_y + s^2 k_x^{'2})^2}{k^{'2}}    e^{-  (\tilde{r}^{-4/3}s k^{'2} +\tilde{r}^{-2/3}s^2k'_x k'_y + s^3 k_x^{'2}/3) }.
\label{Delta2:trans}
\end{eqnarray}
From this expression, we find that $\bar{\Delta}_2(\tilde{r})$
satisfies eq. (\ref{Delta2:long}) and the integrations in eq. (\ref{Delta2:trans})
are reduced to  a constant for $\tilde{r} \gg 1$.
By using a parallel procedure to that for eq. (\ref{Delta2:long}), we find
\begin{eqnarray}
\bar{\Delta}_1(\tilde{r})  & \propto & \tilde{r}^{-11/3}, \qquad \tilde{r} \gg 1, \nonumber \\
\bar{\Delta}_3(\tilde{r})  & \propto & \tilde{r}^{-5/3}, \qquad \tilde{r} \gg 1.
\label{Delta13:trans}
\end{eqnarray}
Substituting eqs. (\ref{Delta2:long}) and  (\ref{Delta13:trans})
into eq. (\ref{Result}), 
we find the long-range correlation in $\bar{C}_{pp}(r)$
as
\begin{eqnarray}
\bar{C}_{pp}(r)  & \propto & \frac{ \rho_0 T_0}{l_c^3} \left ( \frac{r}{l_c} \right )^{-5/3}, \qquad r \gg l_c. \label{bar_C}
\end{eqnarray}
Thus, we confirm that the threshold length $l_c$ in eq. (\ref{threshold})
plays an important role.

\subsection{Numerical Simulation}

% Setting

To verify our theoretical prediction, 
we perform simulation of three-dimensional hard spherical particles.
In our simulation $m$, $\sigma$ and the average temperature $T_0$ are set to be unity, 
and all quantities are converted to dimensionless forms,
where the unit of time scale is $\sigma \sqrt{m/T_0}$.
We adopt the Lees-Edwards boundary condition \cite{Evans}.
The volume fraction $\nu$ is
$\nu_c/8$ with the closest packing fraction of 
particle $\nu_c$.
We use the parameters  $\dot{\gamma}=0.5$ and $e=0.83$
to keep the temperature unity.
We examine the size of the system $L= 28, 56, 112$ in our simulation,
which contain $4096$, $32768$, and $262144$ particles, respectively.

%\subsection{Velocity Auto-correlation function}

\begin{figure}
  \includegraphics[height=.3\textheight]{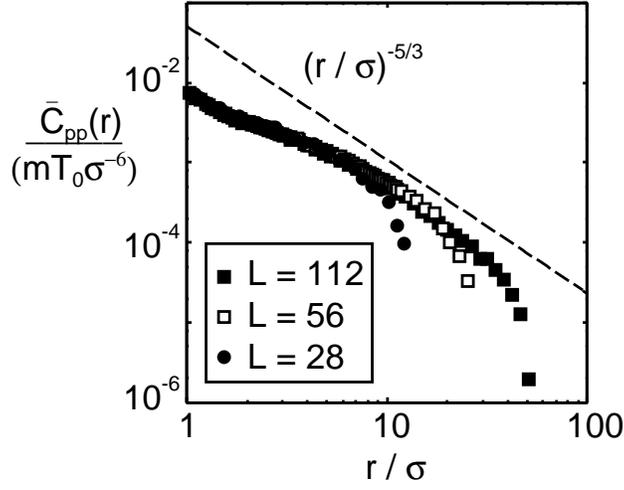}
  \caption{The angular average $\bar{C}_{pp}(r)$ of the momentum density
for $L = 28, 56, 112$ as a function of the distance $r$.}
  \label{C_p}
\end{figure}

Figure \ref{C_p} shows the numerical result of 
$\bar{C}_{pp}(r)$ 
for various system size $L$.
There is an apparent finite size effect,
where $\bar{C}_{pp}(r)$ decays faster than power-law function for $r > 0.3L$.
For the system with the largest size, 
we find the existence of a region where the spatial correlation function
$\bar{C}_{pp}(r)$ approximately satisfies $r^{-5/3}$.
We cannot confirm the existence of the region to obey $r^{-5/3}$
in the wide range, but the results seem to be
consistent with our theoretical result in eq. (\ref{bar_C}).

\section{Conclusion and Discussion}

% Validity of the analysis

Let us discuss our results. From the condition $l_c^{-2} \sim O(k^2)$,
we could obtain the analytic expression of $C_{pp}(\bv{r})$.
This condition can be rewritten  as $l_c \sim O(L/(2 \pi))$.
For the case of our numerical simulation,
$l_c$ is estimated as $1.6\sigma$. 
On the other hand, $L/(2 \pi)$ is estimated as $18 \sigma$ in the largest system
in our numerical simulation.
Hence, it might be suspicious that the condition  $l_c \sim O(L/(2 \pi))$
is satisfied in our numerical simulation.
However, the result of the simulation seems to be consistent with our analytic
result.
We need more careful consideration on the validity of our analytic method. 

It should be noted that our method in not valid in the true hydrodynamic limit.
Indeed, we have assumed eq. (\ref{condition}), which can be rewritten as 
$k \sigma \sim \epsilon^{1/4}$.
Thus, if $\epsilon$ is finite, the applicable range of our analysis is limited
for smaller systems.
We may expect different features in the region of $k \sigma \ll \epsilon^{1/4}$,
which will be our future task.

% Correlation function in dense system

The theoretical method we have used is applicable only to the dilute system.
In general, sheared granular systems in experiments are not dilute ones.
 Hence, it is not clear that the long-range correlation
predicted in this paper is experimentally observable.
In order to theoretically understand the spatial correlation in
such dense sheared granular systems, we must improve our theoretical method,
which will be also our future work.

% Conclusion

In conclusion, we have analytically calculated the behavior of
$C_{pp}(\bv{r})$ in three-dimensional dilute sheared granular fluids. 
Based on fluctuating hydrodynamics, we find that
there is a long-range correlation in
$C_{pp}(\bv{r})$.
The results are verified by numerical simulations.

\begin{theacknowledgments}
We thank H. Wada and V. Kumaran for valuable discussion.
This work is partially supported by Ministry of Education, Culture, Science and Technology (MEXT), Japan (Grant No. 18540371) and the Grant-in-Aid for the global COE program
"The Next Generation of Physics, Spun from Universality and Emergence"
from the Ministry of Education, Culture, Sports, Science and
Technology (MEXT) of Japan.
One of the authors (M. O.) thanks the Yukawa Foundation for financial
support.
The numerical calculations were carried out on Altix3700 BX2 at YITP in Kyoto University.
\end{theacknowledgments}

\appendix
\section{APPENDIX : the elements of the matrix $L^{(i)}_{\alpha \beta}$
and the vector $R_\alpha(\bv{k},t)$}

In this appendix, we explicitly show the elements of the matrix $L^{(i)}_{\alpha \beta}$
and the vector $R_\alpha(\bv{k},t)$.
Here, we non-dimensionalize the variables by
the mass $m$, the characteristic time $\nu_0 = \rho_0 T_0 / (m\eta_0(T_0))$,
and the characteristic length $l_0 = 2 \sqrt{T_0/m} / \nu_0$.
The matrix $L^{(i)}$ and the vector $R_\alpha(\bv{k},t)$ 
are explicitly given by
\begin{eqnarray}
L^{(1)}_{\alpha \beta} & = & 
\left[ 
\begin{array}{ccccc}
0 & 0 & 1 & 0 & 0\\
0 & 0 & \sqrt{2/3} & 0 & 0 \\
1 & \sqrt{2/3}& 0 & 0 & 0 \\
0 & 0 & 0 & 0 & 0 \\
0 & 0 & 0 & 0 & 0 \\
\end{array} 
\right],
\end{eqnarray}
\begin{eqnarray}
L^{(2)}_{\alpha \beta} & = & 
\left[ 
\begin{array}{ccccc}
0 & 0 & 0 & 0 & 0\\
5\sqrt{3}\mu^*/(4\sqrt{2})  &  5\lambda^*/4 & 0 & 0 & 0 \\
0 & 0 & 2 \eta^*/3 & 0 & 0  \\
0 & 0 & 0 & \eta^*/2 & 0\\
0 & 0 & 0 & 0 & \eta^*/2 \\
\end{array} 
\right],
\end{eqnarray}
\begin{eqnarray}
L^{(3)}_{\alpha \beta} & = & 
\left[ 
\begin{array}{ccccc}
0 & 0 & 0 & 0 & 0\\
0 & 0 & 0 & 0 & 0 \\
0 & 0 & k_x k_y/k^2 & 2 k_x k_\perp/k^2 & 0  \\
0 & 0 & -k_x/k_\perp  & -k_x k_y/k^2 & 0  \\
0 & 0 & -k_y k_z/k k_\perp  & -k_z/k & 0  \\
\end{array} 
\right],
\end{eqnarray}
and 
\begin{eqnarray}
\tilde{R}_1 & = & 0, \nonumber \\
\tilde{R}_2 & = & - \sqrt{3/2}\alpha_h i k_i \tilde{f}^{h}_i
- \sqrt{2/3}\alpha _s \dot{\gamma} \tilde{f}^{s}_{xy} /\nu_{0}, \nonumber \\
\tilde{R}_{2+l} & = & \alpha_s i e^{(l)}_i k_j \tilde{f}^{s}_{ij}, 
\end{eqnarray}
where $k_\perp^2 = k^2 - k_y^2$, $\alpha_s =  \{ m\eta^*/(\rho_0 l_0^3) \}^{1/2} $,
and $\alpha_h =  \{ 5m\kappa^* / (3\rho_0 l_0^3)\}$.
$\tilde{f}^{h}_i$ and $\tilde{f}^{s}_{ij}$
are the Fourier transforms of $f^h_i$ and $f^s_{ij}$, respectively.
We do not describe the explicit form of $L^{(4)}$ and $L^{(5)}$ in eq. (\ref{matrix})
because they affect only the higher order corrections in eq. (\ref{Result}).

%%%%%%%%%%%%%%%%%%%%%%%%%%%%%%%%%%%%%%%%%%%%%%%%
%% BACKMATTER
%%%%%%%%%%%%%%%%%%%%%%%%%%%%%%%%%%%%%%%%%%%%%%%%



\bibliographystyle{aipproc}   % if natbib is available

\bibliography{sample}

\IfFileExists{\jobname.bbl}{}
 {\typeout{}
  \typeout{******************************************}
  \typeout{** Please run "bibtex \jobname" to optain}
  \typeout{** the bibliography and then re-run LaTeX}
  \typeout{** twice to fix the references!}
  \typeout{******************************************}
  \typeout{}
 }

%%%%%%%%%%%%%%%%%%%%%%%%%%%%%%%%%%%%%%%%%%%
%% The following lines show an example how to produce a bibliography
%% without the help of the BibTeX program. This could be used instead
%% of the above.
%%%%%%%%%%%%%%%%%%%%%%%%%%%%%%%%%%%%%%%%%%%

\end{document}